\documentclass[journal, preprint]{vgtc}
\ieeedoi{10.1109/TVCG.2026.3747431}

\onlineid{0}

\vgtccategory{Research}

\title{Having Dog Ears ``for Real'': Effects of Active and Passive \\ Haptics on Embodying Non-Human Body Parts in VR}

\author{%
  \authororcid{Omar A. Khan}{0009-0000-1705-0950},
  \authororcid{Bao Han Trinh}{0009-0006-5112-1912},
  \authororcid{Lee Lisle}{0000-0002-7252-5987}, and
  \authororcid{Tiffany D. Do}{0000-0003-3323-4586}
}

\authorfooter{
  \item
  	Omar A. Khan is with Drexel University.
  	E-mail: oak44@drexel.edu
  \item
  	Bao Han Trinh is with Drexel University.
  	E-mail: bt596@drexel.edu
  \item
  	Lee Lisle is with Virginia Tech.
  	E-mail: llisle@vt.edu
  \item
  	Tiffany D. Do is with Drexel University.
  	E-mail: td848@drexel.edu
}

\abstract{%
Embodying non-human body parts in VR is a prevalent practice among certain subcultures and is a personally important creative outlet to many individuals. However, the discrepant morphology between real and virtual bodies can decrease Sense of Embodiment (SoE). Haptic feedback can compensate by increasing SoE felt towards non-human body parts, but there is a literature gap in comparing the effects of different haptic modalities, and their combinations, on SoE. Through an online survey sent out to social VR communities ($n=63$), we determined that animal ears are a commonly embodied and ecologically valid non-human body part to study. We then ran a $2\times2$ within-subjects user study ($n=28$) with two independent variables: \textit{active haptics}, delivered through vibrotactile gloves, and \textit{passive haptics}, delivered through a physical headband, for when participants reach up to touch virtual dog ears appended to their avatar in VR. Our findings show that (1) passive haptics produced the strongest overall embodiment outcomes, (2) combining modalities reduced the benefits of passive haptics, and (3) SoE towards non-human body parts positively correlates with SoE towards the entire avatar.
} %


\teaser{
  \centering
  \includegraphics[width=\linewidth]{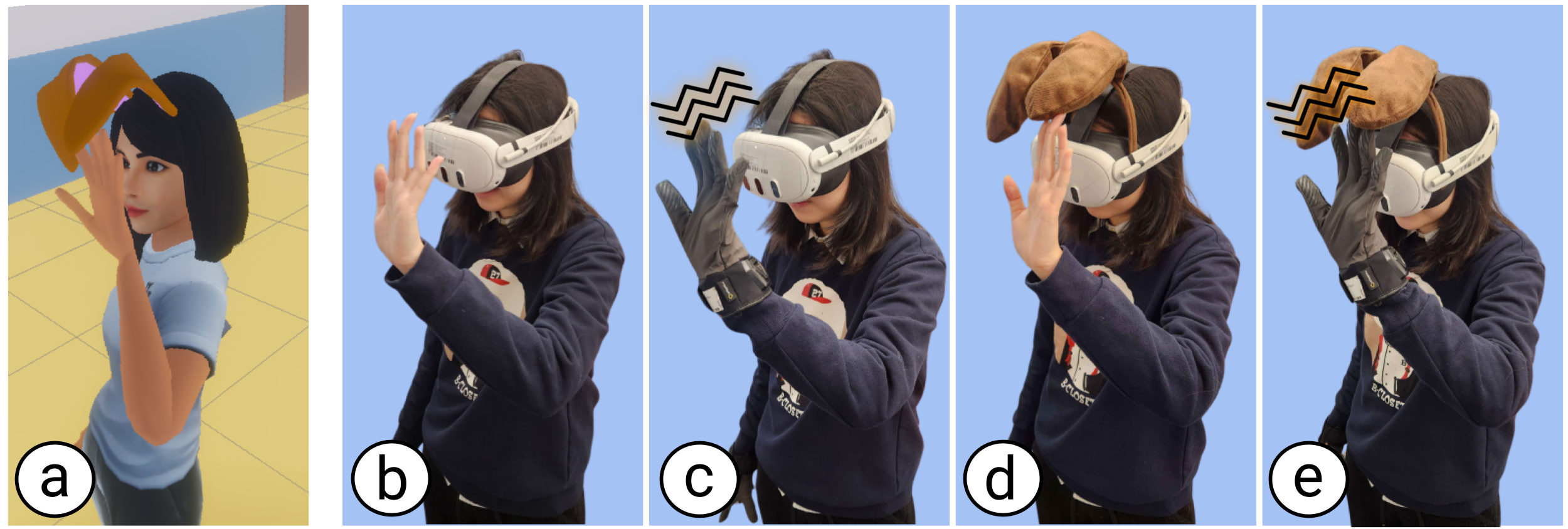}
  \vspace{-1.2em}
  \caption{An overview of our user study, including the haptic combinations: (a) an embodied humanoid avatar with dog ears appended to its head, (b) ActOff-PasOff: no haptics, (c) ActOn-PasOff: only active haptics, (d) ActOff-PasOn: only passive haptics, (e) ActOn-PasOn: both active and passive haptics.}
  \label{fig:teaser}
}

\usepackage{booktabs}      
\usepackage{soul}
\usepackage{mathptmx}      

\begin{document}

\firstsection{Introduction}
\maketitle

\label{sec:intro}

In many virtual reality (VR) applications, users are represented by an avatar, a virtual body which they embody, control, and often deliberately shape to express themselves. Social VR platforms, such as VRChat, allow users to create and customize their own avatars, offering a powerful tool for self-expression. Prior work shows that people have used their avatars in social VR to explore personal interests and skills~\cite{freeman_my_2020} and to approach diverse gender identities~\cite{freeman_rediscovering_2022}. Social VR users have also associated VRChat experiences with improvements in mental health and well-being, especially in improving social connection and reducing isolation~\cite{deighan_social_2023}. For example, autistic individuals have used their avatars in social VR as a means of expressing themselves authentically, without the fear of judgment and social constraints of the physical world~\cite{grillo_exploratory_2025}.

Within these environments, many users choose avatars that diverge from human morphology by incorporating non-human appendages such as animal ears, tails, and wings~\cite{takashita_escape_2026}. This practice is particularly prominent in furry and kemonomimi avatars, which account for a substantial portion of the players in social VR applications~\cite{dong_exploring_2024}.
Furry avatars are donned by members of the furry fandom (furries), a subculture centered on interest in anthropomorphic animals. Furries often represent themselves through personalized anthropomorphic animal characters known as fursonas, both within social VR~\cite{dong_exploring_2024, grillo_exploratory_2025} as well as at real-world events and conventions dating back to the 1980s~\cite{patten_furry_2017}.
Kemonomimi, a Japanese term originating from anime and manga communities, refers to humanoid characters with animal ears and sometimes a tail, and these avatars are likewise frequently embodied in social VR~\cite{dong_exploring_2024, shijo_research_2022}.

A challenge which arises while embodying avatars with non-human body parts is that the discrepancy between users' real and virtual bodies can affect Sense of Embodiment (SoE)~\cite{vargas_now_2023, do_stepping_2024}, a term which refers to the sensations of being inside, having, and controlling a virtual body~\cite{guy_sense_2023}. SoE is central to facilitating the aforementioned benefits of embodied social VR experiences~\cite{freeman_my_2020}.
Body ownership is a component of SoE~\cite{roth_construction_2020}, and it refers to the extent to which users feel as though their avatar is their own body.
Body ownership plays a key role in shaping how users behave and interact in virtual environments~\cite{banakou_illusory_2013}, and it is highly important to social VR users~\cite{freeman_my_2020}.
A promising solution to mitigate the discrepancy between users' real and virtual bodies is to induce a body ownership illusion (when users experience their actual body being replaced by their virtual body~\cite{mottelson_systematic_2023}), through the integration of tactile and haptic feedback~\cite{shi_dobbyear_2025, botvinick_rubber_1998}.
Haptic feedback has been shown to enhance SoE both towards avatars as a whole~\cite{dewez_towards_2021}, as well as towards non-human appendages~\cite{chang_hapticwings_2025, egeberg_extending_2016}.

VR haptics can be subdivided into \textit{active haptics} and \textit{passive haptics}~\cite{vaghela_active_2021, fouad_touching_2025}. Active haptics are provided by computer-controlled devices (e.g. vibrotactile responses or force feedback), whereas passive haptics are provided by static physical props which are positionally synced with objects in the virtual environment.
Prior research has found that passive haptics outperform active haptics for emotional bonding when petting a virtual dog~\cite{fouad_touching_2025}, and motor performance in a tapping task~\cite{mcanally_visualhaptic_2022}.

There have also been work on hybrid techniques which integrate various haptic modalities, including work by Sun et al.~\cite{sun_smart_2019}, Achibet et al,~\cite{achibet_flexifingers_2017}, and Lee et al.~\cite{lee_multimodal_2024}. This direction is especially relevant to non-human embodiment, since many users in the furry fandom integrate wearable props, such as fursuits~\cite{patten_furry_2017} into embodied experiences, suggesting that physical props are already meaningful in this design space. Since these experiences often already include vibrotactile feedback, it is important to examine how passive and active haptics work together and impact SoE.

To the best of our knowledge, there is no comparative study on how active haptics and passive haptics impact SoE towards non-human body parts appended to a user's avatar, or what the effect of combining those modalities would be.
This knowledge can be highly important to social VR users who adopt avatars with non-human or extra appendages, such as furry avatars and kemonomimi avatars.
For example, since passive haptics via static props are often cheaper and more accessible than active haptic devices, they may be used to save costs if their impact on SoE is not significantly different across the two modalities.
Additionally, when researching SoE towards individual body parts, it is useful to consider its relationship to SoE towards the entire avatar as part of an interconnected system. This relationship is currently underexplored in the literature, and it is an angle we aim to explore in the current work.

To investigate the research gaps described above, we devised the following research questions (RQs):

\begin{itemize}[leftmargin=6.3ex]
\setlength\itemsep{0pt}
    \item [\textbf{RQ1.}] How do haptic modalities (active vs passive) differentially affect users’ SoE towards non-human body parts?
    \item [\textbf{RQ2.}] How does combining active and passive haptics affect SoE towards non-human body parts?
    \item [\textbf{RQ3.}] What is the relationship between SoE towards non-human body parts, and SoE towards the whole avatar?
\end{itemize}

For the remainder of this paper, we review related work and present two studies we conducted in order to address the above RQs.
The first study is a formative survey we ran with frequent VRChat users regarding common avatars with non-human body parts to ensure ecological validity of our results.
The survey gave us insights into which non-human body parts are most commonly adopted in social VR.
We used insights from our survey to guide the development and design of our second study, which is a controlled $2\times2$ within-subjects experiment in which users used active and passive haptic interfaces and reported their effects on SoE towards their virtual dog ears, and towards their whole avatar. Our findings contribute to literature on embodiment and VR haptics, and have implications in various domains.

\section{Related Work}

The current work builds on prior research in embodiment perception, non-human embodiment, and haptic feedback in VR. In this section, we review related work in each of those topics.

\subsection{Embodiment Perception}

Embodiment refers to our conscious awareness of our bodies~\cite{chang_hapticwings_2025}.
Early research on embodiment in real environments has shown that, through the assistance of synchronized tactile feedback, users perceive their actual body being replaced by external objects~\cite{armel_projecting_2003}, a phenomenon known as the body ownership illusion~\cite{mottelson_systematic_2023}.
An early example is the rubber hand illusion discovered by Botvinick et al.~\cite{botvinick_rubber_1998}, in which body ownership is induced to a rubber hand being stroked synchronously with the user's concealed real hand.

Unlike embodiment in real environments, virtual embodiment is mediated by avatars, and has been a topic of particular interest in VR research.
Slater et al.~\cite{slater_inducing_2009} brought the concept of body ownership to VR, showing that visuo-haptic synchrony (synchronized touch between real and virtual bodies) induces body ownership.
This concept was reinforced in studies by Normand et al.~\cite{normand_multisensory_2011} and Kilteni et al.~\cite{kilteni_extending_2012}.
Visuo-motor synchrony (syncing movement of the avatar with real body movements) has likewise been shown to be central to inducing a SoE~\cite{kokkinara_measuring_2014, pfeiffer_multisensory_2013} in VR.
An interesting emerging phenomenon concerning embodiment, especially with frequent social VR users~\cite{chen_understanding_2024} is phantom touch, which is when users experience a sense of touch in VR without any real-life contact or haptics~\cite{alexdottir_phantom_2022, alexdottir_phantom_2025}.
Prior research has also explored how avatar appearance~\cite{do_stepping_2024, do_cultural_2024, waltemate_impact_2018} and perspective~\cite{slater_first_2010, maselli_building_2013} affect SoE in VR.
Various standardized questionnaires have been constructed to measure SoE, such as the Standardized Embodiment Questionnaire (SEQ) by Peck and Gonzales-Franco~\cite{peck_avatar_2021}, and the Virtual Embodiment Questionnaire (VEQ) by
Roth and Latoschik~\cite{roth_construction_2020}, which argues that SoE is a fundamental attribute of self-consciousness that arises from multi-sensory information processing, including visual and haptic stimuli.

Embodiment in avatars holds special importance to frequent social VR users~\cite{freeman_my_2020}. Commercially available social VR applications such as VRChat allow users to customize and embody their own avatars, which has been applied in social VR settings to attain a variety of positive outcomes, such as enhanced social support~\cite{li_we_2023, deighan_social_2023}, recreation~\cite{maloney_stay_2021}, intimacy in long-distance relationships~\cite{maloney_falling_2020, freeman_hugging_2021}, and exploration of self-identity~\cite{freeman_my_2020, freeman_rediscovering_2022}.
As social VR is rising in popularity, especially with children and teens~\cite{maloney_stay_2021, hinduja_metaverse_2024}, the study of embodiment, and factors that shape SoE (such as haptic feedback), are becoming increasingly crucial and relevant.
Despite this, there remain notable gaps in embodiment literature, such as the distinction and relationship between SoE towards individual body parts, and SoE towards the entire avatar, which we aim to cover in the current work.
Embodiment of non-human body parts is also a research area that is highly deserving of attention, as we discuss in the following subsection.

\subsection{Non-Human Embodiment}

In VR, avatars are not limited to the standard human form.
The malleable nature of avatar embodiment allows users to experiment with their digital self-representations.
In social VR applications such as VRChat, users often prefer to have their avatars reflect their idealized self~\cite{freeman_my_2020}, instead of having a `look-alike' avatar.
In particular, embodying non-human body parts is a common practice~\cite{takashita_plausible_2026} that is personally important to those who are in certain subcultures, such as furry or anime subcultures, as it allows them to embody non-human characters that they identify with (e.g. fursonas, kemonomimi)~\cite{grillo_exploratory_2025, dong_exploring_2024}.
These characters incorporate animal body parts to various degrees, such as animal ears and tails~\cite{shijo_research_2022}. We explore the non-human body parts embodied by real VRChat users in our field survey in Section~\ref{sec:survey}.

SoE, and body ownership illusion, can be induced towards animal and animal-like avatars~\cite{krekhov_vr_2018, lan_can_2023}.
Lanier~\cite{lanier_homuncular} introduced the concept of homuncular flexibility, which is the concept that humans can learn to control bodies with non-human morphology and accept them as their own bodies.
Similarly to human body parts, visuo-motor synchrony enhances SoE towards non-human body parts such as tails~\cite{steptoe_human_2013, ito_we_2019}.
A prominent theme in research involving non-human embodiment is to enhance nature connectedness~\cite{ahn_experiencing_2016, spangenberger_becoming_2022} and empathy towards animals and endangered species. Past research has shown that embodying animals can enhance empathy felt towards beavers~\cite{sierra_rativa_can_nodate}, sea turtles~\cite{pimentel_effects_2022}, penguins~\cite{gagnon_waddle_2023}, and stray cats and dogs~\cite{xu_istraypaws_2024}.
Other than empathy, embodying non-human avatars has also been shown to change other attitudes and behaviors of users, such as reducing fear of height~\cite{oyanagi_transformation_2019}, improving acting performance~\cite{kammerlander_using_2021}, and influencing locomotion preferences~\cite{khan_i_2025, khan_impact_2025}.
Non-human embodiment enables beyond-human locomotion capabilities~\cite{tanahashi_experiencing_2026}.
Past research has also explored control mappings to embody avatars with non-human morphology, such as the work by Krekhov et al.~\cite{krekhov_vr_2018,krekhov_beyond_2019}, and HandAvatars by Jiang et al.~\cite{jiang_handavatar_2023}.
Thayer et al.~\cite{thayer_investigating_2025} evaluated different control mappings for embodying spiders in VR, and Takashita et al.~\cite{takashita_embodied_2024} evaluated different finger mappings to control a 12-jointed tentacle appended to the user's avatar.
Egan et al.~\cite{egan_dog_2024} used a deep-learning method to map human motion to quadreped motion for controlling quadrupeds while embodying them.
Recently, Takashita et al.~\cite{takashita_plausible_2026} proposed a design framework for embodying non-human avatars while maintaining plausible actions and sensations.

A challenge with embodying avatars which diverge from the standard human form, such as those with non-human body parts, is that the morphological differences between the user's real and virtual body may limit SoE~\cite{vargas_now_2023}.
Haptic feedback has shown promise as a solution.
For instance, past studies
found that SoE towards virtual wings appended to a humanoid avatar could be enhanced by vibrotactile feedback~\cite{egeberg_extending_2016}, and 2D weight-shifting~\cite{chang_hapticwings_2025}.
Past research has also shown that synchronous touch enhances SoE when the users are embodied in fully non-human avatars, such as bat avatars~\cite{andreasen_spatial_2018} and dog avatars~\cite{vargas_now_2023}.
Kondoh et al.~\cite{kondoh_move_2025} used personalized force feedback to enhance SoE towards non-human avatars (e.g. ammonites), using Bayesian optimization to adjust haptic parameters to maximize plausibility for each individual user.
Shi et al.~\cite{shi_dobbyear_2025} found that tactile feedback could be elicited for non-human embodiment without using any mechanical devices or proxies, but instead through haptic retargeting (redirecting the movement path of the users' hand when they reach up to touch their ears).
Together, these papers suggest that haptic feedback can enhance SoE for non-human body parts, and there are various methods for delivering effective haptic feedback.
Notable gaps exist in \textit{directly comparing} different haptic modalities in their effectiveness in enhancing SoE towards non-human body parts, as well as the effects of \textit{combining} those modalities. These are key questions driving the current work.

\subsection{Active and Passive Haptic Feedback}

As discussed in previous subsections, haptic feedback is a factor which has repeatedly shown to enhance SoE in VR, towards both human body parts and non-human body parts.
VR haptics can be broadly classified into two categories: active haptics, and passive haptics.
Active haptics are provided by computer-controlled devices. They are often hand-held or wearable, and they render sensations of touch using techniques such as shape displacement~\cite{benko_normaltouch_2016, yoshida_pocopo_2020}, pivoting motors/actuators~\cite{kovacs_haptic_2020, chang_hapticwings_2025, kondoh_move_2025}, and vibrations~\cite{egeberg_extending_2016, fouad_touching_2025}.
The majority of prior studies on the role of haptic feedback in enhancing SoE employs active haptic devices.
The inherent limitation of active haptics is that there is no perceived force which stops the body from moving into virtual objects, which limits realism~\cite{suzuki_hapticbots_2021}.

Alternatively, passive haptics involve the use of physical artifacts as a proxy, so that users can touch and interact with a real object which is positionally synced with a virtual object in VR~\cite{suzuki_hapticbots_2021}.
For example, Hettiarachchi and Wigdor proposed Annexing Reality~\cite{hettiarachchi_annexing_2016}, a system which scans objects in a user's environment and matches the position and shape of virtual objects with them, to provide on-demand passive haptics.
Another technique is Haptic Retargeting~\cite{azmandian_haptic_2016, cheng_sparse_2017}, which involves redirecting the user's hand to the positions of virtual objects, and was recently shown to be effective in enhancing embodiment of non-human body parts~\cite{shi_dobbyear_2025}.
To enhance the applicability of passive proxies, researchers have combined passive haptics with other technologies, such as redirected walking~\cite{kohli_combining_2005}, robotic arms~\cite{araujo_snake_2016, vonach_vrrobot_2017}, shape displays~\cite{iwata_project_2001, nakagaki_inforce_2019}, and mobile robots~\cite{suzuki_hapticbots_2021, he_physhare_2017}.
The advantage of passive haptics is that it can generate veridical haptic and proprioceptive sensations at a low cost, and can stop the body from moving into virtual objects.
However, syncing the position and shape of physical proxies to real objects is a challenge, and a mismatch can break the illusion~\cite{simeone_substitutional_2015}.

Active and passive haptics each present benefits and limitations in VR.
Prior studies have directly compared the two modalities in various interaction scenarios.
In a highly relevant study, McAnally et al.~\cite{mcanally_visualhaptic_2022} found that passive haptics enhanced SoE in a tapping task compared to active haptics, although active haptics yielded greater SoE than the no haptics at all, and combining the two modalities did not present any interaction effects.
However, this study was mainly focused on the task performance, with haptics being provided for the tapping task rather than an interaction with a non-human appendage.
Fouad et al.~\cite{fouad_touching_2025} compared active and passive haptics in a virtual dog interaction, and found similar results on emotional bonding, for which passive haptics outperformed active haptics, which outperformed the baseline.
Together, the above studies show promise that passive haptics \textit{may} outperform active haptics in enhancing SoE towards non-human body parts on a user's avatar in VR, potentially helping save costs for those who adopt and hold importance towards avatars with non-human parts.
However, this has not been empirically tested prior to the current work.

\section{Formative Survey}
\label{sec:survey}

The most prominent real-world use case of non-human embodiment is by certain online subcultures, such as the furry and kemonomimi fandoms, who embody non-human body parts in social VR applications such as VRChat~\cite{dong_exploring_2024}.
To ensure an ecologically valid experimental design, we sent out an online survey to determine which non-human body parts are commonly embodied in VRChat.
This survey was approved by the Drexel University Institutional Review Board (Protocol Number: 2511011464), and was sent out to VRChat communities through social media channels (Discord and the VRChat forum) and advertised at an in-person convention centered around the furry fandom~\footnote{\url{https://nordicfuzzcon.org/} (Accessed 2026-03-05)} .
To ensure we received responses from our target population, we placed particular focus on recruiting from furry and kemonomimi communities.

\begin{table}[h!]

\centering
\caption{The top ten non-human body part categories most commonly embodied in VRChat, based on our online survey ($n=63$).}
\footnotesize
\label{tab:survey}
\begin{tabular}{l|c}
\toprule
\textbf{Body part category} & \textbf{Count} \\
\midrule
Tails & 57 \\
\hl{Animal/elf ears} & \hl{51} \\
Snout/muzzle & 45 \\
Animal extremities (paws/claws/hooves/talons) & 35 \\
Horns & 27 \\
Fur/mane/fluff & 27 \\
Animal legs (digitigrade/quadruped) & 25 \\
Wings & 23 \\
Scales & 8 \\
Fangs & 7 \\
\bottomrule
\end{tabular}
\end{table}

Through our online survey, 63 participants listed the non-human body parts they frequently embody while using VRChat, and the non-human body parts they often see other users embodying while using VRChat. We aggregated the body parts into categories and counted the number of participants which mentioned each category. The top ten most commonly mentioned body part categories are listed in Table~\ref{tab:survey}.
We also collected demographics data on the sample. The complete survey responses are included in our supplementary materials.

For our in-person user study, we opted to use \textbf{floppy dog ears},
which belongs to the second most prominent body part category based on our survey (highlighted in Table~\ref{tab:survey}).
They also fit well within our technical limitations, as the hand-tracking on the Quest 3 loses accuracy directly above the user’s head. Floppy ears would droop down in front of the HMD and help minimize the effects of tracking degradation.
As 51 out of 63 responses mentioned animal/elf ears, including two responses which specifically mentioned floppy animal ears, we consider floppy dog ears to be an appropriate choice for our in-person user study.

\section{Experiment}

This section details the in-person VR user study we ran on our university campus to answer our RQs.

\subsection{Materials}

\begin{figure}[h]
    \vspace{-0.5em}
    \centering
    \includegraphics[width=0.7\linewidth]{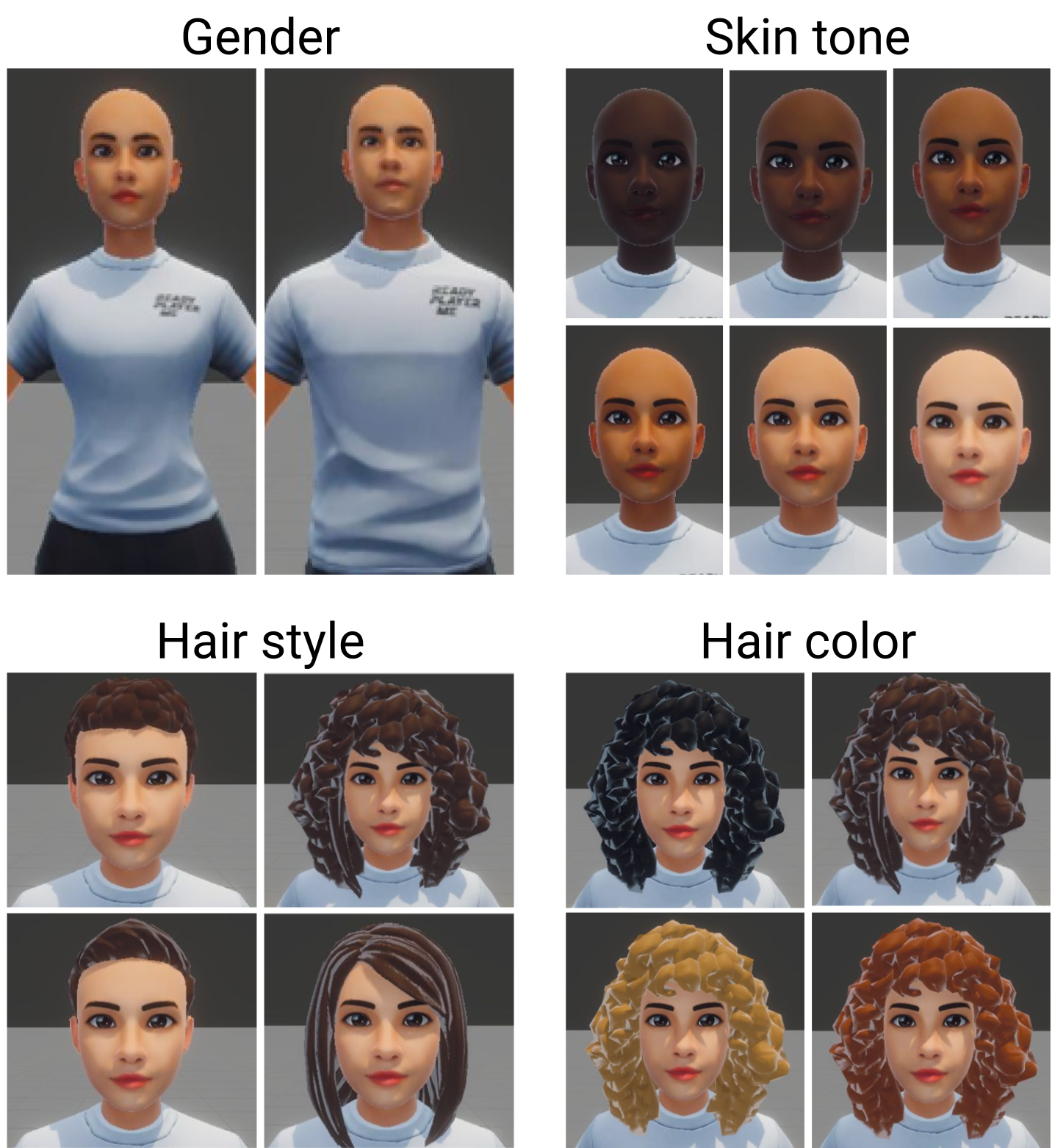}
    \vspace{-0.5em}
    \caption{Avatar customization options.}
    \label{fig:customization}
\end{figure}

\paragraph{\textbf{Avatar Implementation}}
We used the Unity game engine (v6000.0.39f1) to develop a VR application in which users would embody avatars which were exported from the ReadyPlayerMe avatar creation platform.
Users were able to customize the avatar according to a preset set of traits for gender, skin color, hair style, and hair color (see Figure~\ref{fig:customization}).

The avatar embodiment and tracking was implemented using inside-out hand tracking from the built-in sensors on the Meta Quest 3 head-mounted display.
Following the implementation of Khan et al.~\cite{khan_influence_2026}, the animation of the limbs and torso of the avatars were realized using the FinalIK API, following the ReadyPlayerMe Unity integration guide.
We used the inverse kinematics settings provided by the guide. The height of the avatars were individually calibrated based on the position of the headset while worn by the participants.
Participants were positioned 1.5 meters away from a virtual mirror to give them a view of their avatar for the duration of the experiment.

\begin{figure}[h!]
    \centering
    \includegraphics[width=0.7\linewidth]{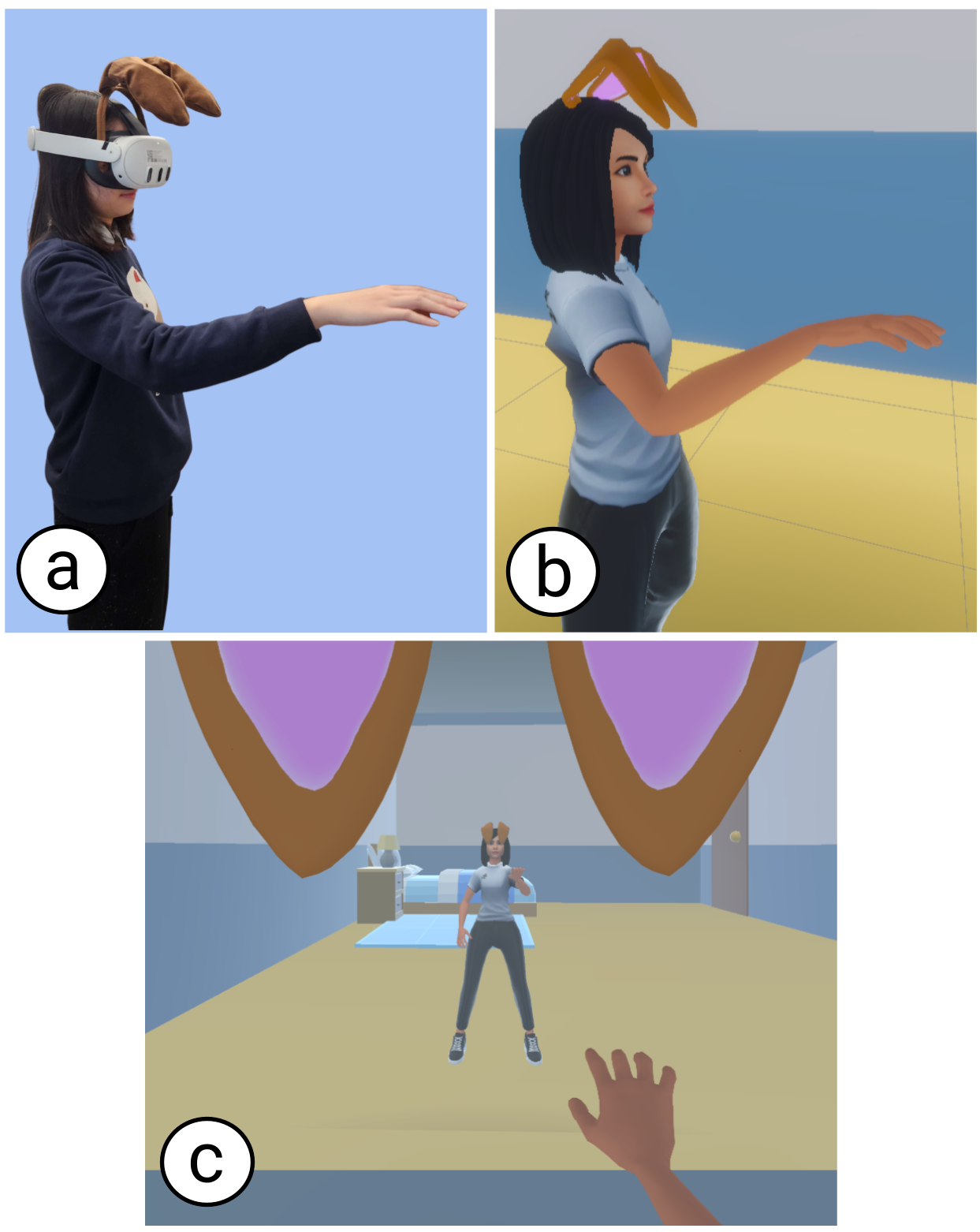}
    \vspace{-0.5em}
    \caption{The embodiment and first-person perspective of participants during the experiment.}
    \label{fig:embodiment}
\end{figure}

We appended a 3D model of floppy dog ears to the avatar's head. The dog ears droop down in front of the avatar's face, causing their tips to be visible from the first-person perspective of the participant, as shown in Figure~\ref{fig:embodiment}a. This configuration is similar to the dog ear accessories commonly used by players in social VR environments.
The left and right ears in the 3D model each contained four bones. We applied physics and gravity to the bones closest to the tips, and enabled collision between the ears and the avatar hands to allow for realistic movement and manipulation of the dog ears.

\begin{figure}[h]
    \vspace{-0.5em}
    \centering
    \includegraphics[width=0.7\linewidth]{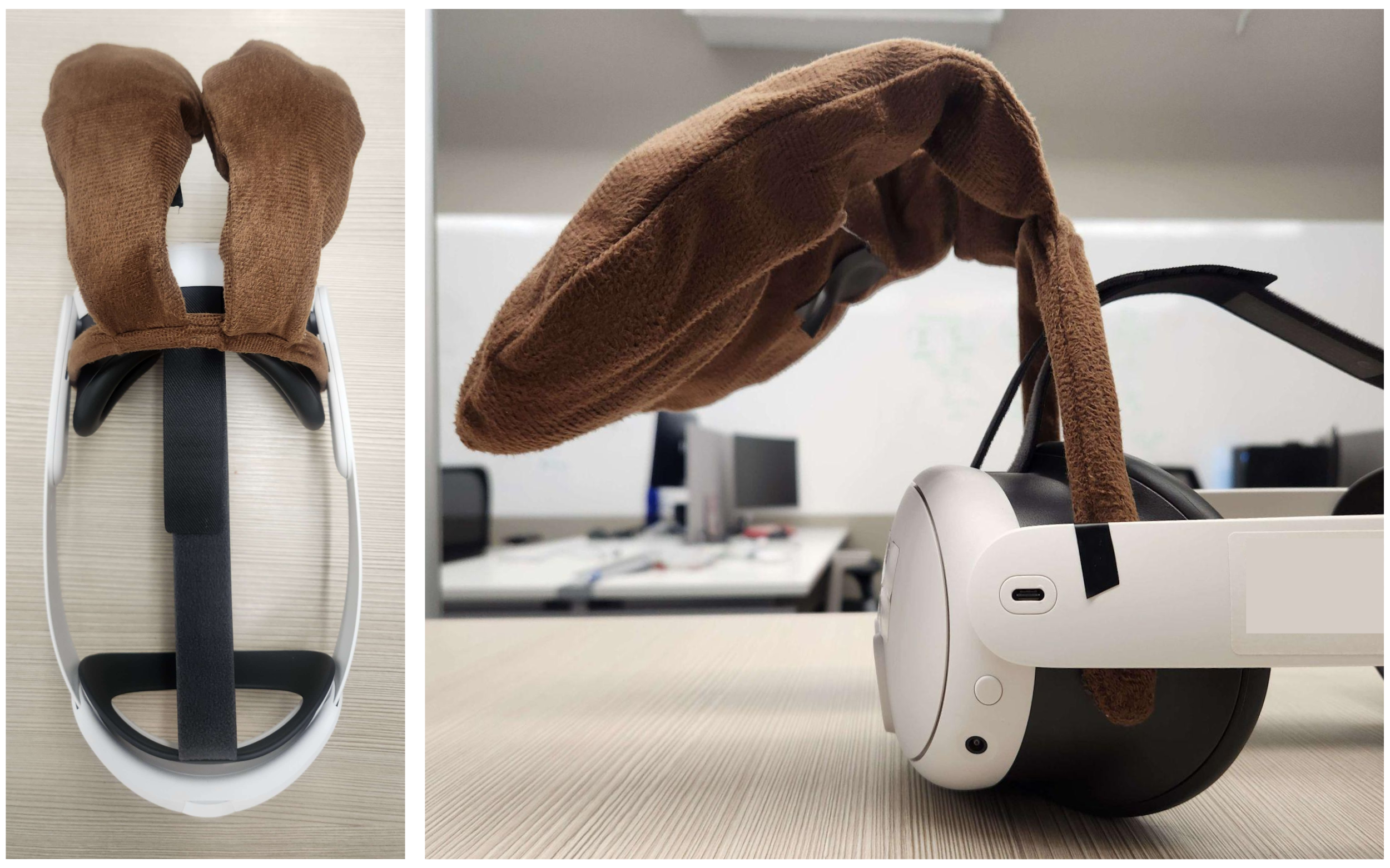}
    \vspace{-0.5em}
    \caption{The dog ear headband which was mounted to the HMD and used as a physical proxy in the PasOn conditions.}
    \label{fig:headband}
\end{figure}

\paragraph{\textbf{Active Haptics Implementation}}

Following previous research on active haptics in VR~\cite{fouad_touching_2025}, we used the bHaptics TactGloves DK2~\footnote{BHaptics TactGloves DK2: \url{https://www.bhaptics.com/en/tactsuit/tactglove-dk2/} (accessed 2026-03-05)} to deliver active haptics in the form of vibrotactile feedback.
The gloves contain vibration actuators on each fingertip and the palm, enabling the design of flexible time-controlled haptic patterns that respond to user input. Due to this flexibility, the device has been widely used in VR haptics research.

Within the Unity scene, colliders were placed on the fingertips and the palms of each hand, as well as on the virtual dog ears. When a participant's virtual hand contacted the virtual dog ears in VR, the colliders would intersect, sending a signal to the haptic gloves to vibrate the intersection point for 0.10 seconds. Using the bHaptics designer, we set the vibration intensity to 20\%, matching the method of a relevant prior user study~\cite{fouad_touching_2025}. These parameters provided localized vibrotactile contact cues, with feedback delivered at the level of individual fingers.

\paragraph{\textbf{Passive Haptics Implementation}}

Passive haptics in VR are generated by physical proxies in the real world that correspond to the position and shape of virtual objects.
Therefore, to generate passive haptics for the virtual dog ears, we used a dog ear headband which we mounted to the head-mounted display, as shown in Figure~\ref{fig:headband}.

The surface material of the headband was polyester and the ears were stuffed with cotton to approximate the texture and softness of dog ears. The headband matched the shape of the virtual dog ears, and was positionally aligned with them. To ensure consistent alignment across study sessions and conditions, black tape markers were placed on the sides of the head mounted display to standardize the position and orientation of the headband, as shown in Figure~\ref{fig:headband}. When users reached up to touch their virtual dog ears, they would feel the physical ears of the headband, and were able to feel and manipulate it in sync with the virtual dog ears, generating realistic passive haptic sensations.

\subsection{Independent Variables}

Our user study implemented a $2\times2$ within-subjects design, with two independent variables (active haptic feedback and passive haptic feedback), and two levels for each (off and on), resulting in four experimental conditions:
\begin{itemize}[leftmargin=3ex]
\setlength\itemsep{0pt}
    \item \textbf{ActOff-PasOff}: No haptics. Participants interacted with their virtual dog ears without using any haptic feedback.
    \item \textbf{ActOn-PasOff}: Active haptics only. Participants interacted with their virtual dog ears with vibrotactile feedback provided by haptic gloves.
    \item \textbf{ActOff-PasOn}: Passive haptics only. Participants interacted with their virtual dog ears while wearing a dog ear headband as a physical proxy.
    \item \textbf{ActOn-PasOn}: Both active and passive haptics. Participants wore both haptic gloves and the dog ear headband concurrently.
\end{itemize}
We removed the haptic gloves in Act-Off conditions to better isolate the effect of active haptics and more closely approximate typical social VR interaction without haptic gloves.
Participants completed all four conditions in a counterbalanced order based on a Latin square design, for which 24 orderings were used.

\subsection{Dependent Variables}
\label{sec:measures}

We adopted the Virtual Embodiment Questionnaire (VEQ) by Roth and Latoschik~\cite{roth_construction_2020}
as it has been previously used to measure SoE of animal body parts~\cite{vargas_now_2023, krekhov_beyond_2019}.
Our purpose was to measure participants' SoE towards both their virtual dog ears and the overall avatar, for each of the haptic combinations.
The VEQ measures three factors of virtual embodiment: ownership, agency, and change.
Of these, we decided to adopt ownership and agency, as we thought these were the most suitable for the purposes of our study, and in-line with prior research on non-human embodiment~\cite{vargas_now_2023}.
To address RQ3, we measured both the ownership of the dog ears as well as ownership of the entire avatar.
Since our study does not provide a control mechanism for the virtual dog ears, we did not measure agency towards the dog ears, instead opting to use the original questions provided in the VEQ for that factor. Our questions for each measure are provided in Appendix~\ref{appendix:likert}.

Since we modified the body ownership questions from what was originally present in the VEQ, we computed Cronbach's alphas for ownership of the whole body and ownership of the ears, which each demonstrated good internal consistency with values of $\alpha=.882$ and $\alpha=.853$ respectively.
In addition to our embodiment questionnaire, we also collected written responses to open-ended questions (see Appendix~\ref{appendix:open-ended}) at the end of each experimental session, for participants to provide reasoning for their responses.

\subsection{Hypotheses}

Based on prior work on active and passive haptic feedback in VR, we devised the following hypotheses for our RQs. \textbf{H1a} and \textbf{H1b} correlate with \textbf{RQ1}, \textbf{H2} correlates with \textbf{RQ2}, and \textbf{H3} correlates with \textbf{RQ3}.

\begin{itemize}[leftmargin=6.3ex]
\setlength\itemsep{0pt}
    \item [\textbf{H1a}] Passive haptics will outperform active haptics in enhancing SoE.
    \item [\textbf{H1b}] Active haptics will compensate in enhancing SoE when passive haptics are absent.

    \item [\textbf{H2.}] Combining passive and active haptic feedback will decrease SoE compared to single-modality conditions.

    \item [\textbf{H3.}] Sense of ownership and agency towards the whole avatar will be positively correlated with sense of ownership of the virtual dog ears.
\end{itemize}

\textbf{H1a} and \textbf{H1b} are based on the findings of Fouad et al.~\cite{fouad_touching_2025}, from a study on active vs passive haptics in a virtual dog interaction.
\textbf{H2} is based on prior research that suggests that combining feedback modalities can increase workload~\cite{vitense_multimodal_2003}, which can degrade the sense of agency~\cite{jahanian-najafabadi_task_2025}, a component of SoE.
\textbf{H3} is based on prior research on body ownership illusion, in which users perceived non-human body parts as parts of their real body~\cite{shi_dobbyear_2025}.

\subsection{Participants}
\label{sec:participants}
We recruited 28 participants through flyers and listservs.
In a pre-screening survey, the participants provided their consent to participate, and were pre-screened to satisfy our recruitment criteria, such as normal or corrected-to-normal vision, the ability to move both arms freely, and no severe history of cybersickness.
To ensure diversity in our sample, we collected participant demographics in our pre-study screening survey, including age ($M=23.8$, $SD=4.9$, $range=19-36$), gender (man: 17, woman: 10, agender: 1), and sexual orientation (straight: 22, bisexual: 3, demisexual: 1, queer: 1, prefer not to say: 1).

We also asked participants to report how often they use VR, for which 11 said ``never", 9 said ``once or twice ever", 2 said ``once or twice per month", 4 said ``once or twice per week" and 1 said ``three to four times per week".
Since our study was focused on embodying a personalized avatar, we also asked participants to report how often they use a personalized avatar, for which 9 said ``never", 8 said ``once or twice ever", 2 said ``once or twice a year", 3 said ``once or twice a month", 1 said ``once or twice per week", 2 said ``three to four times per week", and 3 said ``daily/almost daily".

\subsection{Procedure}

The study consisted of one in-person session that lasted approximately
45 minutes and received ethics approval by the Drexel University Institutional Review Board (Protocol Number: 2511011464). Participants first completed an online screening survey that captured
their demographics, described in Section~\ref{sec:participants}.

Upon arrival to the study room, participants were assigned to one of four
Latin squares ordering cohorts for counterbalancing the four within-subject conditions.
Subsequently, they were instructed to customize their avatar's character model according to a preset set of traits for gender, skin color, hair style, and hair color (see Figure~\ref{fig:customization}). Prior research has shown that users feel greater SoE while embodying avatars which match their gender and race~\cite{do_stepping_2024}. Therefore, we advised participants to choose the traits they felt corresponded to their real life image the most.
\begin{table*}[t]
    \caption{Means and standard deviations for all measurements.}
    \label{tab:summary}
    \centering

    \begin{tabular}{l|cccc}
         \toprule
            & \textbf{ActOff-PasOff} & \textbf{ActOn-PasOff} & \textbf{ActOff-PasOn} & \textbf{ActOn-PasOn} \\
        \midrule
             Ownership of Whole Body&  5.063 (1.395)&  5.107 (1.091)&  5.366 (1.075)&  4.938 (1.152) \\
             Ownership of Dog Ears&  3.631 (1.472)&  4.262 (1.486)&  5.048 (1.381)&  4.523 (1.511)  \\
             Agency&  5.446 (.941)&  5.339 (.960)&  5.777 (.911)&  5.098 (1.155) \\
         \bottomrule
    \end{tabular}
\end{table*}

Following the avatar selection, participants donned the VR headset, along with either haptic gloves, physical headband, both, or none, adhering to the Latin squares ordering assigned to them. They
were then instructed to stand straight in order to calibrate the avatar’s
height (using FinalIK’s calibration), with the avatar remaining concealed from their view until after calibration. The FinalIK
calibration system scales the avatar to match the height of the headset.
Subsequently, participants listened to audio instructions and engaged in a standard embodiment procedure, adapted from the protocol developed by Roth and Latoschik~\cite{roth_construction_2020} (see Figure~\ref{fig:embodiment}a-b).
During these tasks, participants executed a series of actions in
front of a virtual mirror (see Appendix~\ref{appendixB}).
Since our questionnaire involved questions on SoE felt towards virtual dog ears appended to the avatar, we included an additional set of ten instructions pertaining to interaction with the virtual dog ears.

After the final task, participants exited VR and participants filled out a questionnaire on a computer. This process was then repeated for the remaining three conditions.
After the last condition, participants typed responses to open-ended questions.
Our full embodiment questionnaire and open-ended questionnaire are provided in Appendix~\ref{appendix:questionnaires}.
Participants were then thanked for their time and were compensated with a gift card.%

\section{Results}

\begin{figure}[t]
    \vspace{-.2em}
    \centering
    \includegraphics[width=\linewidth]{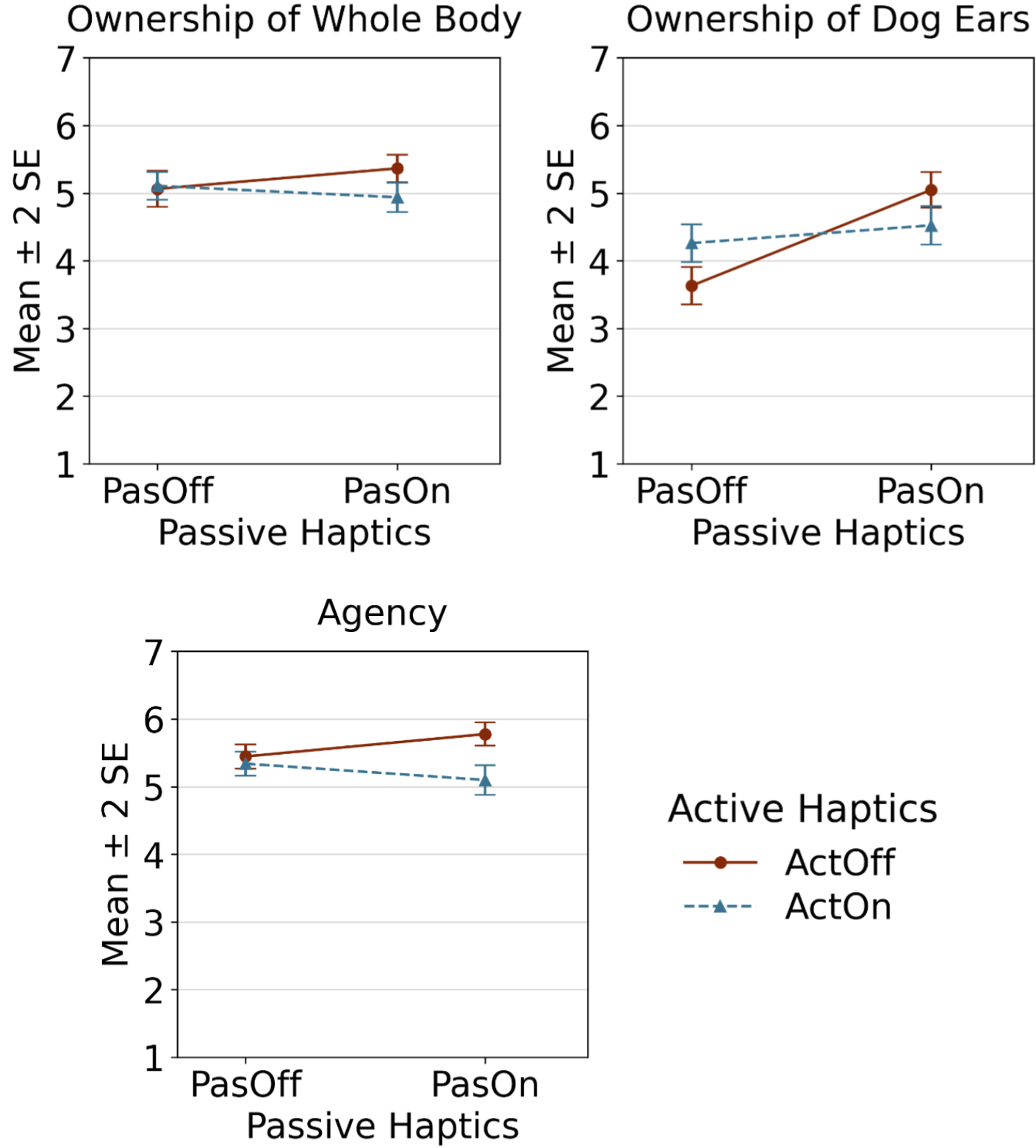}
    \vspace{-1.2em}
    \caption{Interaction plots visualizing how combinations of active and passive haptics affect SoE. All measures were assessed on a 7-point Likert scale.}
    \label{fig:two-way}
    \vspace{-1em}
\end{figure}

In this section, we present the results of our experiment, which we obtained by analyzing the questionnaire data using IBM SPSS Statistics (v31.0.0.0).
We conducted a two-way repeated measures ANOVA test at a 95\% confidence level, with two independent variables (active and passive) and two levels for each (off and on).
Figure~\ref{fig:two-way} presents interaction plots to visualize our results.

\begin{figure}[t]
    \vspace{-.4em}
    \centering
    \includegraphics[width=\linewidth]{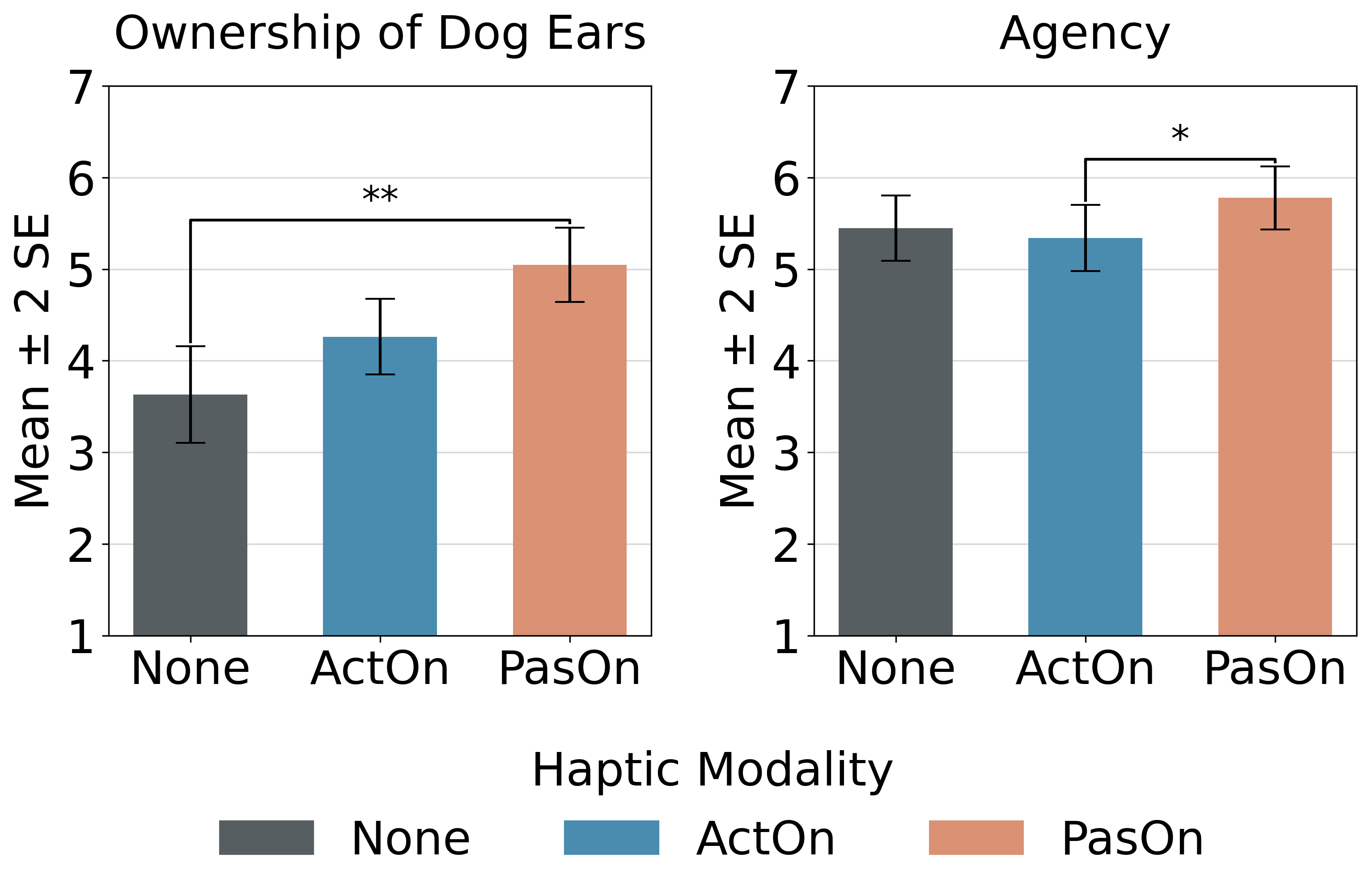}
    \vspace{-1.2em}
    \caption{Visualized comparisons of single-modality conditions. All measures were assessed on a 7-point Likert scale. Asterisks denote a significant difference between conditions (*: $p<.05$, **: $p<.01$).}
    \label{fig:one-way}
    \vspace{-1em}
\end{figure}

Due to the presence of interaction effects for all of our measures in the two-way ANOVA test, we could not interpret the main effects in isolation. We therefore conducted simple effects analyses to interpret the interaction.
To address RQ1, which asks whether active or passive haptics are more effective when only one modality is available, we conducted an omnibus one-way repeated-measures ANOVA comparing the three single-modality conditions: none (ActOff-PasOff), active-only (ActOn-PasOff), and passive-only (ActOff-PasOn).
Figure~\ref{fig:one-way} visualizes differences between these single-modality conditions.
For significant omnibus effects, we then report the planned linear contrast across these ordered conditions, followed by Bonferroni-adjusted pairwise comparisons used to interpret the effect.

To account for multiplicity across our primary hypothesis tests, we additionally applied the Benjamini-Hochberg false discovery rate (FDR) procedure across the family of primary tests corresponding to H1-H3: the three interaction effects from the 2 × 2 repeated-measures ANOVAs, the three omnibus one-way repeated-measures ANOVAs comparing the single-modality conditions, and the three correlation analyses. Simple effects, planned linear contrasts, and pairwise comparisons were treated as follow-up analyses and were not included in this FDR family. We report FDR-adjusted p-values for the primary tests discussed below.
All of our primary conclusions are robust to correction for multiple comparisons.

All of the assumptions for a repeated measures ANOVA tests were met, including the same subjects being present in all groups, no significant outliers, and Maunchly's test of sphericity.
Bonferroni corrections were applied to the post hoc comparisons, and the p-values reported below are Bonferroni-adjusted values.
To address \textbf{RQ3}, we conducted Pearson correlation tests, which we have also detailed in this section.
The full questionnaire data for all participants is included in our supplementary materials.

\subsection{Ownership of Whole Body}

\paragraph{\textbf{Interaction effects}}
Our two-way ANOVA revealed an interaction effect ($F(1, 27) = 5.589$, \textbf{\itshape{p}\,$=$\,.026}, \textbf{\itshape{pFDR}\,$=$\,.033}, $\eta^2_p = .171$) between active haptics and passive haptics, on sense of ownership towards the whole body.
Simple effects analyses revealed that when passive feedback was on (PasOn), the presence of active feedback \textit{reduced} sense of ownership towards the whole body (ActOff-PasOn\,$>$\,ActOn-PasOn, \textbf{\itshape{p}\,$=$\,.006}).

\paragraph{\textbf{Comparison of single-modality conditions}} A one-way repeated-measures ANOVA comparing the none (ActOff–PasOff), active-only (ActOn–PasOff), and passive-only (ActOff–PasOn) conditions did not reveal a significant effect of haptic modality on ownership of the whole body.

\subsection{Ownership of Dog Ears}

\paragraph{\textbf{Interaction effects}}
Our two-way ANOVA revealed an interaction effect ($F(1, 27) = 9.809$, \textbf{\itshape{p}\,$=$\,.004}, \textbf{\itshape{pFDR}\,$=$\,.007}, $\eta^2_p = .266$) on sense of ownership towards the dog ears.
Simple effects analyses revealed that when passive haptics were off (PasOff), the presence of active haptics \textit{enhanced} ownership towards dog ears (ActOn-PasOff > ActOff-PasOff, \textbf{\itshape{p}\,$=$\,.023}), but when passive haptics were on (PasOn), the presence of active haptics \textit{reduced} ownership towards the dog ears (ActOff-PasOn > ActOn-PasOn), \textbf{\itshape{p}\,$=$\,.048}).
Additionally, when active haptics were off (ActOff), the presence of passive haptics \textit{enhanced} ownership towards dog ears (ActOff-PasOn\,$>$\,ActOff-PasOff, \textbf{\itshape{p}\,$<$\,.001}).

\paragraph{\textbf{Comparison of single-modality conditions}}
An omnibus one-way repeated-measures ANOVA comparing the none (ActOff-PasOff), active-only (ActOn-PasOff), and passive-only (ActOff-PasOn) conditions revealed a significant effect of haptic modality on ownership of the dog ears,
($F(2, 54) = 10.250$, \textbf{\itshape{p}\,$<$\,.001}, \textbf{\itshape{pFDR}\,$\leq$\,.002}, $\eta^2_p = .275$). The planned linear contrast was also significant, ($F(1, 27) = 16.427$, \textbf{\itshape{p}\,$<$\,.001}, $\eta^2_p = .378$). Pairwise comparisons revealed that passive haptics enhanced sense of ownership towards the dog ears, compared to no haptics (ActOff-PasOn\,$>$\,ActOff-PasOff, \textbf{\itshape{p}\,$=$\,.001}).
Additionally, we observed a trend of passive haptics outperforming active haptics (ActOff-PasOn\,$>$\,ActOn-PasOff, \textbf{\itshape{p}\,$=$\,.066}), although the p-value for this trend did not reach our chosen significance threshold.

\subsection{Agency}

\paragraph{\textbf{Interaction effects}}

Our two-way ANOVA revealed an interaction effect ($F(1, 27) = 5.240$, \textbf{\itshape{p}\,$=$\,.030}, \textbf{\itshape{pFDR}\,$=$\,.034}, $\eta^2_p = .163$) between active haptics and passive haptics on sense of agency.
Simple effects analyses revealed that when passive haptics were on (PasOn), the presence of active haptics reduced sense of agency (ActOff-PasOn > ActOn-PasOn, \textbf{\itshape{p}\,$=$\,.004}). Additionally, when active haptics were off (ActOff), the presence of passive haptics enhanced sense of agency (ActOff-PasOn\,$>$\,ActOff-PasOff, \textbf{\itshape{p}\,$=$\,.022}).

\paragraph{\textbf{Comparison of single-modality conditions}}
An omnibus one-way repeated-measures ANOVA comparing the none (ActOff-PasOff), active-only (ActOn-PasOff), and passive-only (ActOff-PasOn) conditions revealed a significant effect of haptic modality on agency, ($F(2, 54) = 4.202$, \textbf{\itshape{p}\,$=$\,.020}, \textbf{\itshape{pFDR}\,$=$\,.030}, $\eta^2_p = .135$).
The planned linear contrast was also significant, ($F(1, 27) = 5.944$, \textbf{\itshape{p}\,$=$\,.022}, $\eta^2_p = .180$). Pairwise comparisons revealed that passive haptics enhanced sense of agency compared to active haptics (PasOn-ActOff\,$>$\,PasOff-ActOn, \textbf{\textit{p\,$=$\,.016}}).

\subsection{Correlations}

\begin{figure}[h!]
    \vspace{.5em}
    \centering
    \includegraphics[width=0.5\linewidth]{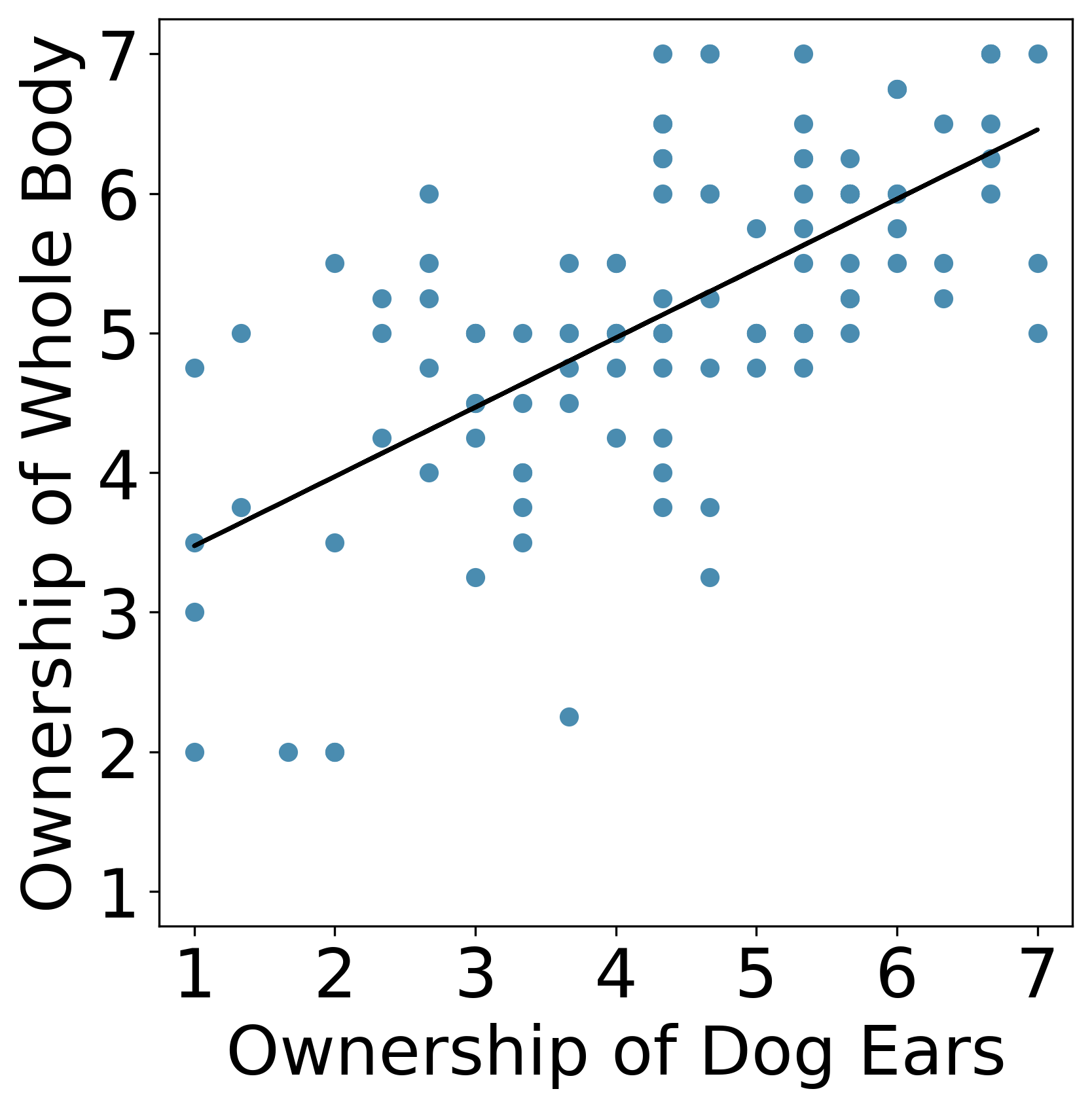}
    \caption{Positive correlation between Ownership of Dog Ears and Ownership of Whole Body ($r=.651$, $p<.001$, $pFDR\leq.002$, $n=28$)}
    \label{fig:correlation}
    \vspace{.5em}
\end{figure}

To determine the relationship between SoE towards non-human body parts and SoE towards the entire avatar body (\textbf{RQ3}), we ran Pearson correlation tests on each of our measures, and found that all three measures are positively correlated with one another. The correlation between Ownership of Dog Ears and Ownership of Whole Body, visualized as a scatterplot in Figure~\ref{fig:correlation}, is particularly compelling and relevant. Details on each of the correlations are provided in Table~\ref{tab:correlations}.

\begin{table}[h!]
\centering
\caption{Pearson correlations between our measurements.}
\footnotesize
\setlength{\tabcolsep}{4.5pt}
\label{tab:correlations}
\begin{tabular}{l|c|c}
\toprule
\textbf{Variables correlated} & \textbf{$r$} & \textbf{$pFDR$} \\
\midrule
Ownership of Dog Ears $\times$ Ownership of Whole Body & $.651$ & $\leq.002$ \\
Ownership of Dog Ears $\times$ Agency (of Whole Body) & $.484$ & $\leq.002$ \\
Ownership of Whole Body $\times$ Agency (of Whole Body) & $.724$ & $\leq.002$ \\
\bottomrule
\end{tabular}
\\[1ex]
\footnotesize{$N=28$ for all correlations; two-tailed tests.}
\end{table}

\section{Discussion}

In this section, we identify three key findings based on our results, support them with participant quotes, and discuss implications on various domains and on the broader embodiment literature. We also discuss limitations of the current work, and outline future work.

\subsection{Key Findings}

\paragraph{\textbf{[RQ1] Passive haptics produce the strongest overall embodiment outcomes}}

The primary objective of this paper (\textbf{RQ1}) was to compare active and passive haptic feedback to learn which modality has a greater effect on SoE towards non-human body parts.
Upon analyzing the results of our one-way ANOVA, we found that passive haptics significantly enhanced body ownership towards the dog ears compared to the absence of haptics, with a trend suggesting improvements over active haptics as well.
Additionally, passive haptics significantly enhanced sense of agency compared to active haptics.
These effects are visualized in Figure~\ref{fig:one-way}.

Responses to our open-ended questionnaire provide reasoning for these effects.
Participants felt enhanced realism from the sensations of physically manipulating the ears on the headband.
P6 mentioned that the physical headband gave, \textit{``a real feel while grabbing it and pushing it''}.
P1 stated that they were, \textit{``able to bat them around, and they moved more realistically"}, providing a plausible explanation for the increased sense of agency, compared to the active-only condition.
Participants emphasized the physicality of the headband, which was not present in the no-haptics or active-only conditions.
P17 mentioned, \textit{``I knew there was something physical to perceive and therefore touch"}, while
P15 explained, \textit{``I could actually feel dog ears in front of my face it made me think they were part of my body and I owned them"}.
P21 highlighted the weight of the headband: \textit{``the physical prop added physical weight to my interactions which helps my brain connect with what's being registered in my eyes, rather than just pushing air"}.

These effects show that passive haptics generally produce stronger embodiment outcomes than active haptics, supporting one of our hypotheses, \textbf{H1a}, and building on prior research comparing the effects of passive and active haptics in VR.
Past studies found that passive haptics generally improved user experience compared to active haptics in various interaction scenarios within VR, such as emotional bonding~\cite{fouad_touching_2025} and motor performance~\cite{mcanally_visualhaptic_2022}. This finding confirms that this phenomenon extends to the domain of embodiment, expands our understanding of the inter-relation of embodiment and haptics, and contributes to embodiment literature, especially concerning non-human embodiment.

However, \textbf{H1b} was rejected, as we did not observe a significant improvement of active haptics over the no-haptics condition.
H1b was based on the work of Fouad et al.~\cite{fouad_touching_2025}, who found that active haptics could compensate for the absence of passive haptics during interaction with an external virtual dog.
Rejection of H1b suggests that active haptics may be effective for conveying responsiveness from a virtual companion, but less effective for establishing ownership of an additional body part, which may depend more strongly on body-relevant multisensory consistency and physical correspondence.

Passive haptic props are significantly cheaper than active haptic devices, and
it is notable that a cheap haptic interface resulted in stronger overall embodiment compared to a costlier haptic interface.
These finding can help reduce costs in recreational contexts, such as social VR users seeking to enhance embodiment of non-human parts (e.g. those in the furry and kemonomimi communities).
The implications of our findings may also extend to more serious domains, such as frontier exploration or teleoperation, although such applications are speculative in nature.

\paragraph{\textbf{[RQ2] Combining modalities reduces the benefits of passive haptics}}

The secondary objective of this paper (\textbf{RQ2}) was to investigate the effects of combining active and passive haptic feedback on SoE towards non-human body parts. Our two-way ANOVA revealed significant interaction effects of active haptics and passive haptics on Ownership of Whole Body, Ownership of Dog Ears, and Agency. Analyzing the simple effects revealed that, for all three of our measures, combining modalities reduced SoE compared to passive haptics only. These effects are visualized in Figure~\ref{fig:two-way}.

According to responses from our open-ended questionnaire, the most common justification for the decreased SoE resulting from combining the modalities was \textbf{sensory overload}. P9 mentioned that using both haptic modalities was, \textit{``over stimulating and the fealing [sic] of realism was gone"}. P26 elaborated, \textit{``it creates too many senses at the same time, both real and the effects from the gloves, the feeling of my ownership becomes much more fake"}. P25 echoed this sentiment: ``\textit{the feedback from the gloves provided more stimulation than I could process in tandem with attempting to feel the prop through the glove"}. P19 stated, \textit{``the combination hurt my ownership sense. I would reach for the ears, but now expecting two responses, the vibration of the haptic gloves and the softness of the physical ears, which made me actually more focused on the fact that I was interacting with two different worlds rather than just the virtual"}. Similarly, P7 found the combination to be, \textit{``more distracting"}.
Other than sensory overload, other justifications included, \textit{``if a physical object is present, there isn't much need for a virtual sense of feeling, as the physical object can already be felt"} (P22), and, \textit{``having the gloves on muted the actual feeling of the dog ears"} (P23).

These results support our hypothesis, \textbf{H2}.
Prior research suggests that combining feedback modalities can increase workload~\cite{vitense_multimodal_2003}, which can degrade components of SoE~\cite{jahanian-najafabadi_task_2025}.
The participant quotes align with this argument.
Similarly to the previously discussed observation, this finding may be applied to recreational and serious contexts. It is useful for social VR and gaming, and may also be relevant in broader use cases such as training and teleoperation, although this link is speculative in nature. It suggests that users should not mix haptic modalities if they want to maintain SoE towards extra non-human body parts.

\paragraph{\textbf{[RQ3] SoE towards dog ears is positively correlated with SoE towards the whole body}}

The tertiary objective of this paper (\textbf{RQ3}) was to examine the relationship between SoE felt towards extra non-human body parts, and SoE felt towards the entire avatar as a whole. This is an aspect which we found to be underexplored in embodiment research, and its implications make it an interesting direction to explore. Our correlation analysis revealed that Ownership of Dog Ears was positively correlated with both Ownership of Whole Body and Agency (of the Whole Body), supporting our hypothesis, \textbf{H3}, and suggesting that SoE felt towards non-human parts correlates with general SoE towards the whole body.
This correlation is also supported by responses from our open-ended questionnaire. P22 stated, \textit{``The greater the ownership of the virtual dog ears, the more in sync I felt with the avatar"}. P25 elaborated: \textit{``I felt like the ownership of virtual dog ears contributed positively towards my avatar in a sense that I became more aware of what my avatar is doing"}.

This finding is particularly interesting because the participants' real bodies did not have dog ears, implying that other non-human parts can be appended to user's virtual bodies, and SoE towards those parts will correlate with SoE towards the entire body. This could be useful in various domains which may demand the user to adopt non-human body parts. For example, a user teleoperating a space exploration robot by embodying it in VR could adopt robotic components as parts of their virtual body. In this case, tasks requiring high SoE of robotic components, such as manipulating objects with robotic limbs, could be enabled by enhancing overall SoE experienced by the user.

\subsection{Limitations and Future Work}
A limitation of our work is that, due to our reliance on internal tracking from the head-mounted display, our haptic interfaces may have reduced the quality of the hand tracking, providing a possible alternative explanation of the combined-modality results. BHaptics advertises the TactGloves DK2 to have hand-tracking 90\% as accurate as bare hand tracking~\cite{bHaptics}. Additionally, the dog ear headband may have reduced the range and quality of hand-tracking by partially blocking the cameras of the Quest 3. In future studies, more consideration should be given to maintain consistent hand-tracking quality across all experimental conditions.
Another limitation of the current paper is a gender imbalance in our recruited sample (17 men, 10 women, and 1 non-binary). Although we do not foresee gender differences influencing the results of our experiment, it is important to consider recruiting a sample as balanced as possible to ensure equal and inclusive representation.
Another limitation of our approach is that active and passive haptic conditions differed not only in modality but also in device placement, which may have introduced secondary effects.
We also recruited a largely VR-novice participant sample which may not generalize to the frequent VR users who form the target population of this research.
We also did not include a baseline condition with no dog ears, so we cannot separate the effect of having no dog ears at all from the effect of haptic feedback on full-body ownership.

Future work includes expanding the range of haptic modalities included, and evaluating their individual impacts on SoE. The current work employed vibrotactile haptic gloves as the active haptic interface, which are commonly used in VR studies and serves as a reasonable baseline comparison. However, as haptic rendering technology advances, more advanced types of active haptic interfaces, such as mid-air ultrasound haptics~\cite{hosoi_demonstration_2025}, epidermal haptics~\cite{nittala_like_2019}, and robotic haptics~\cite{suzuki_hapticbots_2021}, should be evaluated for their impact on SoE.
Other non-human body parts could also be considered. We used animal ears due to their prevalence in social VR. However, they limit the applicability of our findings to more serious contexts. Future work could explore robotic or machine-like body parts, and explore whether there are differences in results based on the appearance and context of the non-human body parts used.

\section{Conclusion}

We first ran an online survey to learn the non-human body parts which are commonly embodied in VRChat, through which we identified floppy dog ears as an ecologically valid non-human body part to study. Then, we ran a within-subjects VR experiment to determine the effects of active and passive haptic feedback on SoE towards dog ears attached to the participants' avatar. We found that passive haptics produced stronger overall embodiment outcomes compared to active haptics, and that combining the two modalities reduced the benefits of passive haptics, possibly due to sensory overload. We also found that SoE towards the dog ears is positively correlated with SoE towards the whole avatar. Our findings have implications in both recreational and serious domains, and build upon the literature on embodiment and VR haptics.

\bibliographystyle{abbrv}

\section*{Supplemental Materials}

All supplemental materials are anonymously available on OSF at~\url{https://osf.io/67efr/overview?view_only=86b51ffac34e4f8baac881caf0e936b7}, released under a CC BY 4.0 license. They include (1) responses to our formative online survey, (2) responses to our embodiment questionnaire for the in-person experiment, and (3) a video detailing the experimental design and showing the embodiment procedure.

\bibliography{paper}

\appendix
\section{Questionnaires}
\label{appendix:questionnaires}

\subsection{Modified Virtual Embodiment Questionnaire}
\label{appendix:likert}
\begin{table}[h!]
\centering
\caption{Our embodiment questionnaire, adapted from the VEQ, with changes highlighted. This was repeated for each condition and was assessed on a 7-point Likert scale.}
\label{tab:questionnaire}
\footnotesize
\setlength{\tabcolsep}{2.5pt}
\begin{tabular}{l}
\toprule
\textbf{Ownership of Whole Body} \\
\midrule
It felt like the virtual body was my body.\\
It felt like the virtual body parts were my body parts. \\
The virtual body felt like a \hl{humanoid} body. \\
It felt like the virtual body belonged to me. \\
\midrule
\textbf{Ownership of Dog Ears} \\
\midrule
It felt like the virtual \hl{dog ears} were my body parts. \\
The virtual \hl{dog ears} felt like \hl{dog ears}. \\
It felt like the virtual \hl{dog ears} belonged to me. \\
\midrule
\textbf{Agency} \\
\midrule
The movements of the virtual body felt like they were my movements. \\
I felt like I was controlling the movements of the virtual body. \\
I felt like I was causing the movements of the virtual body. \\
The movements of the virtual body were in sync with my own movements. \\
\bottomrule
\end{tabular}
\vspace{-0.6em}
\end{table}

\subsection{Open-Ended Questionnaire}
\label{appendix:open-ended}
\begin{itemize}[leftmargin=3ex]
\setlength\itemsep{0pt}
    \item [1.] Did the \textbf{haptic glove} impact your sense of ownership of your virtual dog ears (i.e., sense that the ears belong to you)? If so, describe how and why.

    \item [2.] Did the \textbf{physical prop} impact your sense of ownership of your virtual dog ears (i.e., sense that the ears belong to you)? If so, describe how and why.

    \item [3.] Did \textbf{combining} the haptic glove and physical prop impact your sense of ownership of your virtual dog ears (i.e., sense that the ears belong to you)? If so, describe how and why.

    \item [4.] How were the effects of the haptic glove and physical prop \textbf{different} in altering your sense of ownership of your virtual dog ears (i.e., sense that the ears belong to you)? Was one of them more effective than the other? If so, explain why?

    \item [5.] Was there a \textbf{correlation} between your sense of ownership of your virtual dog ears, and how you felt towards your avatar as a whole? If so, explain why.
\end{itemize}

\section{Embodiment Audio Instructions}
\label{appendixB}
This script is adapted from~\cite{roth_construction_2020}.
\begin{itemize}[leftmargin=3ex]
\setlength\itemsep{0pt}
    \item [1.] Stand relaxed, let your arms hang relaxed.

    \item [2.] You will now receive some audio instructions to follow.

    \item [3.] Please perform only the movements specified in the audio instructions and try not to get distracted during this phase.

    \item [4.] We will now begin with the tasks.

    \item [5.] Hold your right arm straight forward with your palm facing down.

    \item [6.] Look at your right hand.

    \item [7.] Look at the same hand in the reflection.

    \item [8.] Put your arm comfortably back down and look at your reflection.

    \item [9.] Hold your left arm straight forward with your palm facing down.

    \item [10.] Look at your left hand.

    \item [11.] Look at the same hand in the reflection.

    \item [12.] Put your arm comfortably back down and look at your reflection.

    \item [13.] Look at the face of your reflection.

    \item [14.] Look at the eye area of your reflection.

    \item [15.] Look at the mouth area of your reflection.

    \item [16.] Look at your right dog ear in your reflection.

    \item [17.] Look at your right dog ear hanging in front of your face.

    \item [18.] Grab the tip of your right dog ear using your right hand.

    \item [19.] With your palm facing away from you, use your right hand to push the tip of your right dog ear forward, away from your face.

    \item [20.] Put your arm comfortably back down and look at your reflection.

    \item [21.] Look at your left dog ear in your reflection.

    \item [22.] Look at your left dog ear hanging in front of your face.

    \item [23.] Grab the tip of your left dog ear using your left hand.

    \item [24.] With your palm facing away from you, use your left hand to push the tip of your left dog ear forward, away from your face.

    \item [25.] Put your arm comfortably back down.

    \item [26.] Please stand still for a moment. The experimenter will now assist you in removing the VR headset.
\end{itemize}

\end{document}